\documentclass[twocolumn,superscriptaddress,amsmath, amssymb, amsfonts,preprintnumbers,aps,prd,longbibliography,nofootinbib]{revtex4-1}

\usepackage{graphicx}
\usepackage{dcolumn}
\usepackage{bm}
 \usepackage{amstext}
 \usepackage{amssymb}
\usepackage{subcaption}
 \usepackage{amsmath}
 \usepackage{graphicx}
 \usepackage{color}
 \usepackage{bbold}
 \usepackage{delimset} 
\usepackage[colorlinks=true]{hyperref}
\usepackage{orcidlink}

\usetikzlibrary{decorations.pathmorphing}
\usetikzlibrary{decorations.markings}
\usetikzlibrary{positioning, shapes, snakes, arrows}
 \usepackage{tikz-feynman}

\tikzset{ 
	graviton/.style={line width=.8pt, -latex,decorate, decoration={snake, segment length=4pt,amplitude=1.8pt, pre length=.1cm, post length=.25cm}},
	worldline/.style={gray, line width=1pt},
	worldlineBold/.style={black, line width=.6pt},
        background/.style={black,dotted,line width=1pt},
	zUndirected/.style={line width=1pt},
	zParticle/.style={line width=1pt,postaction={decorate},decoration={markings,mark=at position .6 with {\arrow[#1]{latex}}}},
	zParticleF/.style={line width=1pt,postaction={decorate}},
	cscalar/.style={line width=1pt,postaction={decorate},decoration={markings,mark=at position .6 with {\arrow[#1]{latex}}}},
	cscalar2/.style={line width=1pt,postaction={decorate},decoration={markings,mark=at position .8 with {\arrow[#1]{latex}}}},
	photon/.style={line width =.8pt, decorate, decoration={snake, segment length=3pt, amplitude=1.8pt,  pre length=.1cm, post length=.1cm}},
	 mid arrow/.style={postaction={decorate,decoration={
        markings,
        mark=at position .5 with {\arrow[#1]{latex}}}}} ,
        worlddot/.style={dotted, line width=.8pt},
	worlddot2/.style={dotted, line width=1pt}   }

\DeclareFontFamily{OT1}{pzc}{}
\DeclareFontShape{OT1}{pzc}{m}{it}{<-> s * [1.350] pzcmi7t}{}
\DeclareMathAlphabet{\mathpzc}{OT1}{pzc}{m}{it}

\def\d{\mathrm{d}}

\def\dd{\delta}

\def\d{\mathrm{d}}

\def\nn{\nonumber}

\def\h{\tfrac12}

\newcommand{\vev}[1]{\langle #1\rangle}

\newcommand{\be}{\begin{equation}}
\newcommand{\ee}{\end{equation}}
\newcommand{\ba}{\begin{align}}
\newcommand{\ea}{\end{align}}
\ifx\genfrac\sdflkaj\else\fi
\newcommand{\sfrac}[2]{{\textstyle\frac{#1}{#2}}}

\newcommand{\mn}{{\mu\nu}}

\begin{document}
\preprint{HU-EP-24/02-RTG}
\preprint{TUM-HEP-1492/24}

\title{Calabi-Yau periods for black hole scattering in classical general
   relativity}

\author{Albrecht Klemm\,\orcidlink{0000-0001-5499-458X}} 
\email{aoklemm@uni-bonn.de}
\affiliation{%
Bethe Center for Theoretical Physics, Universit\"at Bonn, D-53115 Bonn, Germany
}

\author{Christoph Nega\,\orcidlink{0000-0003-0202-536X}}
\email{c.nega@tum.de} 
\affiliation{%
Physics Department, Technical University of Munich, D-85748 Garching, Germany
}

\author{Benjamin Sauer\,\orcidlink{0000-0002-2071-257X}} 
\email{benjamin.sauer@hu-berlin.de}
\affiliation{%
Institut f\"ur Physik, Humboldt-Universit\"at zu Berlin,
D-10099 Berlin, Germany
}

 \author{Jan Plefka\,\orcidlink{0000-0003-2883-7825}} 
\email{jan.plefka@hu-berlin.de}
\affiliation{%
Institut f\"ur Physik, Humboldt-Universit\"at zu Berlin,
D-10099 Berlin, Germany
}

\begin{abstract}
The high-precision description of black hole scattering in classical general relativity using  the post-Minkowskian (PM) expansion requires  the evaluation of  single-scale Feynman integrals at increasing loop orders. Up to 4PM,  the scattering angle and the impulse  are expressible in terms 
of polylogarithmic functions and Calabi-Yau (CY) two-fold periods. As in QFT,   periods of higher 
dimensional CY $n$-folds are expected at higher PM order. We find at 5PM  in the dissipative leading order self-force sector (5PM-1SF) that the only  
non-polylogarithmic functions are the K3 periods encountered before and the ones  of a new hypergeometric CY three-fold. In the 5PM-2SF sector further CY two- and three-fold  periods appear. Griffiths transversality of the CY period motives allows 
to transform the differential equations for the master integrals into $\epsilon$-factorized form and to solve them 
in terms of a well controlled function space, as we demonstrate in the 5PM-1SF sector.  
\end{abstract}
 
\maketitle 


\section{Introduction}
Today's gravitational wave detectors have observed more than one hundred
mergers of binary  black holes (BHs) or neutron star (NSs) systems \cite{LIGOScientific:2016aoc,LIGOScientific:2017vwq,LIGOScientific:2021djp}.
These observations enable studies of fundamental questions
in gravitational-, astro-, nuclear and fundamental physics.  With the upcoming
third generation of gravitational wave detectors \cite{LISA:2017pwj,Punturo:2010zz,Ballmer:2022uxx} 
the need for highest precision predictions of the emitted gravitational
waveforms from theory has arisen \cite{Purrer:2019jcp}.
To achieve this, perturbative analytical and numerical approaches are being followed: The post-Newtonian \cite{Blanchet:2013haa,Porto:2016pyg,Levi:2018nxp}, the 
post-Minkowskian (PM) \cite{Kosower:2022yvp,Bjerrum-Bohr:2022blt,Buonanno:2022pgc,DiVecchia:2023frv,Jakobsen:2023oow}, the self-force \cite{Mino:1996nk,Poisson:2011nh,Barack:2018yvs,Gralla:2021qaf} 
expansion
 as well as
numerical relativity \cite{Pretorius:2005gq,Boyle:2019kee,Damour:2014afa}.
In a synergistic fashion, the import of perturbative quantum field theory (QFT) technology
has considerably extended  our knowledge -- in particular, in the PM expansion that is closest to the considerations
in particle physics.

In the PM approach, one naturally considers the \emph{scattering} 
of BHs or NSs \cite{Kovacs:1978eu,Westpfahl:1979gu,Bel:1981be,Damour:2017zjx,Hopper:2022rwo}, which are modeled as massive point-particles that interact gravitationally in
the logic of effective field theory, due to the scale separation between the objects'
intrinsic sizes (Schwarzschild or neutron star radius) and their separation
($Gm\ll |b|$)
\cite{Goldberger:2004jt}.
Using this effective worldline approach, the two-body scattering observables -- the impulse (change of momentum), the spin kick and the far field waveform --
have been computed up to high orders in the PM expansion,
including spin and tidal effects
\cite{Kalin:2020mvi,*Kalin:2020fhe,*Kalin:2020lmz,*Dlapa:2021npj,*Dlapa:2021vgp,*Liu:2021zxr,*Mougiakakos:2021ckm,*Riva:2021vnj,*Mougiakakos:2022sic,*Riva:2022fru,*Mogull:2020sak,*Jakobsen:2021smu,*Jakobsen:2021lvp,Jakobsen:2021zvh,Jakobsen:2022fcj,Jakobsen:2022psy,Shi:2021qsb,*Bastianelli:2021nbs,*Comberiati:2022cpm,*Wang:2022ntx,*Ben-Shahar:2023djm,*Bhattacharyya:2024aeq,Dlapa:2022lmu,Dlapa:2023hsl,Jakobsen:2023ndj,Jakobsen:2023hig,Jakobsen:2023pvx}.
Complementary QFT approaches
based on scattering amplitudes
have reached similar precision in the PM expansion~\cite{Neill:2013wsa,*Luna:2017dtq,*Kosower:2018adc,*Cristofoli:2021vyo,*Bjerrum-Bohr:2013bxa,*Bjerrum-Bohr:2018xdl,*Bern:2019nnu,*Bern:2019crd,*Bjerrum-Bohr:2021wwt,*Cheung:2020gyp,*Bjerrum-Bohr:2021din,*DiVecchia:2020ymx,*DiVecchia:2021bdo,*DiVecchia:2021ndb,*DiVecchia:2022piu,*Heissenberg:2022tsn,*Damour:2020tta,*Herrmann:2021tct,*Damgaard:2019lfh,*Damgaard:2019lfh,*Damgaard:2021ipf,*Damgaard:2023vnx,*Aoude:2020onz,*AccettulliHuber:2020dal,*Brandhuber:2021eyq,Bern:2021dqo,Bern:2021yeh,Damgaard:2023ttc}. 

In these computations, the advanced toolbox of 
multi-loop Feynman integrals needs to be applied: Generation of the integrand, 
tensor reduction to scalar Feynman integrals and the systematical reduction to a set of
master integrals (MIs) by advanced integration by parts (IBP)  algorithms. The present state-of-the-art
is at the 4PM ($G^4$) or three-loop level~\cite{Bern:2021dqo,Bern:2021yeh,Dlapa:2022lmu,Dlapa:2023hsl,Jakobsen:2023ndj,Jakobsen:2023hig,Jakobsen:2023pvx,Damgaard:2023ttc}. The appearing Feynman integrals go beyond the polylogarithmic case
at the 4PM order, where quadratic combinations of elliptic integrals appear, in a form that 
identifies them with  periods of a one-parameter K3 family, i.e.~a Calabi-Yau (CY) two-fold, 
parametrized by  $x=\gamma-\sqrt{\gamma^{2}-1}$. Here is $\gamma=v_{1}\cdot v_{2}$ with the incoming velocities 
$v_i$ of the BHs.

Families of CY $n$-fold period motives (CYPM) and their extensions describe typically higher-loop parametric 
Feynman integrals in their leading $\epsilon$-order in dimensional regularization. These special functions are generalizations of extensions 
of algebraic or  elliptic functions appearing already at low-loop order 
which can be seen as CY $n$-fold periods for $n=0,1$. See~\cite{Vanhove:2014wqa,Klemm,Bonisch:2021yfw,Bourjaily:2018yfy} 
for reviews of CY $n$-fold families and their mixed Hodge structure 
in this context. The best studied all loop series with systematic 
CY $n$-fold period identification are the $2d$  $(n+1)$-loop Banana graphs~\cite{Bloch:2016izu,Klemm:2019dbm,Bonisch:2020qmm,Bonisch:2021yfw}, 
and the $2d$ $n$-loop fishnet integrals~\cite{Duhr:2022pch}. The corresponding geometries can be  
realized as singular double covers of a Fano base $B$ branched at two times the canonical class of the base $2 K_B$ which shows that they are CY~\cite{Duhr:2022pch}. 
In this paper, we find similar CY $n$-folds in the BH scattering problem.
They appear in the same realization from the Baikov representation of the Feynman integrals. As in~\cite{Bonisch:2020qmm} in some 
cases we find a better smooth realization as complete intersection CY. The 
transcendental functions that determine the $x$ dependence of the impulse and 
scattering angle in the BH  scattering problem  are characterized by CYPM of the corresponding 
families of CY $n$-folds. The latter describe the solutions of the periods as determined by the flatness of the Gauss-Manin (GM) connection with additional structures such as Griffiths transversality (GT), 
integrality and modularity inherited from the CY geometry~\cite{Klemm,Bonisch:2021yfw}. The 
geometrical GM connection is derived from IBP relations for a suitable basis of MIs or alternatively obtained from 
expansions of special Baikov integrals. The mathematical properties of the CYPM 
are necessary to calculate the physical quantities. In particular, we use GT as an essential feature 
to bring the full IBP differential equations into $\epsilon$-factorized form which is convenient to systematically solve them up to the required $\epsilon$-order.

\section{Worldline quantum field theory} A highly efficient tool to address the PM
expansion of the BH or NS scattering problem is the worldline effective field theory
approach \cite{Kalin:2020mvi,Mogull:2020sak}. The spinless compact objects are modeled as point particles. The action takes the compact form 
\begin{align}\label{eq:action}
S=-\sum_{i=1}^{2}m_{i}\!\int\!\d\tau\bigg[\sfrac{1}{2}g_{\mu\nu}\dot x_i^{\mu}\dot x_i^{\nu}
  \!\bigg] + S_{\rm EH}
\end{align}
using proper time gauge $\dot x_{i}^{2}=1$.
The bulk Einstein-Hilbert action $S_{\rm EH}$ includes a de Donder gauge-fixing term, and we employ dimensional regularization with $d=4-2\epsilon$. In the worldline quantum field theory (WQFT)
approach \cite{Mogull:2020sak,Jakobsen:2021zvh} the fields are expanded about their non-interacting background configurations
\begin{align}
  \begin{aligned}\label{backgroundexp}
    x_i^\mu &= b_i^\mu \!+\! v_i^\mu \tau \!+\! z_i^\mu\,, \, \quad 
    g_{\mu\nu} = \eta_{\mu\nu} +\sqrt{32\pi G}\, h_\mn 
  \end{aligned}
\end{align}
with the worldline deflections $z_i^\mu(\tau)$ and graviton field $h_\mn(x)$.
The background data is given by the impact parameter $b^{\mu}=b_{2}^{\mu}-b_{1}^{\mu}$ and
the incoming velocities $v_{1},v_{2}$.
The fields $z^\mu_i$ and $h_{\mu\nu}$ are integrated out in the path integral.
One needs to use retarded propagators as a consequence of the Schwinger-Keldysh in-in formalism \cite{Jakobsen:2022psy,Kalin:2022hph}.
The WQFT interactions contain the standard bulk
graviton vertices as well as worldline vertices coupling a single graviton to worldline
deflections \cite{Jakobsen:2021zvh,Jakobsen:2023ndj}. 
The WQFT tree-level one-point functions
$\vev{z_{i}^{\mu}(\tau)}$ solve the classical equations of motion \cite{Boulware:1968zz} -- trivializing the classical limit. As a consequence, the impulse of the (say) first BH or NS, $\Delta p_{1}^{\mu}$,
follows from the tree-level one point function 
$\Delta p_{1}^{\mu} = \lim_{\omega\to 0}\omega^{2}\vev{z_{1}^{\mu}(\omega)}$ that is evaluated
in the PM expansion. As the worldline vertices only conserve the total inflowing
energy -- opposed to full four-momentum conservation for the bulk graviton vertices -- 
the WQFT tree-level one-point function gives rise to loop-level Feynman
integrals whose order grows with the PM order: The $n$th PM order yields $(n-1)$-loop
integrals (plus a trivial Fourier transform over the momentum transfer $q$).

\section{The impulse in PM expansion}
The PM expanded impulse, 
$
\Delta p_{1}^{\mu} =\sum_{n=1}^{\infty} G^{n} \Delta p^{(n)\,\mu}
$,
may be further subdivided into contributions of different self-force (SF) sectors
according to the scaling in the masses $m_{1}$ and $m_{2}$. 
%
%
%
%
Concretely, we have at 5PM order
\begin{align}
\Delta p^{(5) \mu}_{i}=  m_{1} m_{2} & \Bigl ( m_{1}^{4} \Delta p^{(5) \mu}_{\text{0SF}}
+  m_{1}^{3} m_{2} \Delta p^{(5) \mu}_{\text{1SF}}   \\
+m_{1}^{2} m_{2}^{2} &\Delta p^{(5) \mu}_{\text{2SF}} 
+ m_{1} m_{2}^{3} \Delta \bar{p}^{(5) \mu}_{\text{1SF}} 
 + m_{2}^{4} \Delta \bar{p}^{(5) \mu}_{\text{0SF}}\Bigr) \nn \, .
\end{align}
In fact, the 0SF contributions
$\Delta {p}^{(5) \mu}_{\text{0SF}}$ and $\Delta \bar{p}^{(5) \mu}_{\text{0SF}}$ are linked to geodesic motion and are, in principle, known to all orders in $G$ \cite{Damgaard:2022jem}. The self-force expansion is a complementary perturbative expansion going beyond geodesic motion
in the mass ratio $m_{1}/m_{2}\ll 1$. Importantly, the SF order grows in steps of two in the PM order, e.g.~the first 1SF term appears at 3PM and the first 2SF at 5PM order.
In this paper, we determine the non-polylogarithmic function space of the
1SF terms up to the 5PM order. Moreover, we comment on the situation in the 2SF sector. The impulse is a four-vector and will be
expressed as a linear combination of the four-vectors $b^{\mu}$ as well as
$v_{1}^{\mu}$ and $v_{2}^{\mu}$.

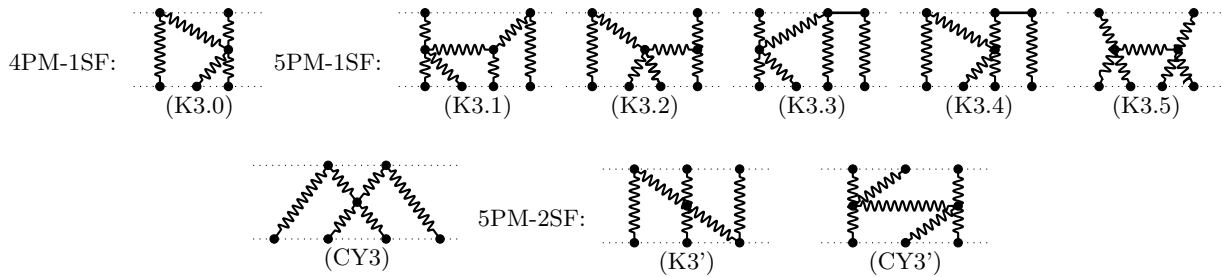
\begin{figure*}[ht!]
4PM-1SF:
\raisebox{-0.9cm}{
\begin{tikzpicture}[baseline={([yshift=-1ex]current bounding box.south)},scale=.7]
 \coordinate (inA) at (0.4,.7);
  \coordinate (outA) at (2.9,.7);
  \coordinate (inB) at (0.4,-.7);
  \coordinate (outB) at (2.9,-.7);
  \coordinate (xA) at (1-0.1,.7);
  \coordinate (xxA) at (1.3,.7);
  \coordinate (yA) at (1.6,.7);
  \coordinate (zA) at (2.2,.7);
  \coordinate (wA) at (2.8+0.1,.7);
  \coordinate (xB) at (1-0.1,-.7);
  \coordinate (xxB) at (1.3,-.7);
  \coordinate (yB) at (1.6,-.7);
  \coordinate (zB) at (2.2,-.7);
  \coordinate (wB) at (2.8+0.1,-.7);
  \coordinate (xM) at (1-0.1,0);
  \coordinate (xxM) at (1.3,0);
  \coordinate (yM) at (1.6,0);
  \coordinate (yyM) at (1.9,0);
  \coordinate (zM) at (2.2,0);
  \coordinate (wM) at (2.8+0.1,0);
  \draw [dotted] (inA) -- (outA);
  \draw [dotted] (inB) -- (outB) node [midway, below] {(K3.0)};
  \draw [photon] (xA) -- (zM);
  \draw [photon] (zM) -- (zB);
  \draw [photon] (xA) -- (xB);
  \draw [photon] (zA) -- (zM);
  \draw [photon] (zM) -- (yB);
  \draw [fill] (xA) circle (.08);
  \draw [fill] (zA) circle (.08);
  \draw [fill] (xB) circle (.08);
  \draw [fill] (yB) circle (.08);
  \draw [fill] (zB) circle (.08);
  \draw [fill] (zM) circle (.08);
\end{tikzpicture}

}5PM-1SF:
\raisebox{-0.9cm}{

\begin{tikzpicture}[baseline={([yshift=-1ex]current bounding box.south)},scale=.7]
 \coordinate (inA) at (0.4,.7);
  \coordinate (outA) at (3.4,.7);
  \coordinate (inB) at (0.4,-.7);
  \coordinate (outB) at (3.4,-.7);
  \coordinate (xA) at (1-0.1,.7);
  \coordinate (xxA) at (1.3,.7);
  \coordinate (yA) at (1.6,.7);
  \coordinate (zA) at (2.2,.7);
  \coordinate (wA) at (2.8+0.1,.7);
  \coordinate (xB) at (1-0.1,-.7);
  \coordinate (xxB) at (1.3,-.7);
  \coordinate (yB) at (1.6,-.7);
  \coordinate (zB) at (2.2,-.7);
  \coordinate (wB) at (2.8+0.1,-.7);
  \coordinate (xM) at (1-0.1,0);
  \coordinate (xxM) at (1.3,0);
  \coordinate (yM) at (1.6,0);
  \coordinate (yyM) at (1.9,0);
  \coordinate (zM) at (2.2,0);
  \coordinate (wM) at (2.8+0.1,0);
  \draw [dotted] (inA) -- (outA);
  \draw [dotted] (inB) -- (outB) node [midway, below] {(K3.1)};;
  \draw [photon] (xA) -- (xM);
  \draw [photon] (xM) -- (xB);
  \draw [photon] (wA) -- (wB);
  \draw [photon] (wA) -- (zM);
  \draw [photon] (xM) -- (yB);
  \draw [photon] (xM) -- (zM);
   \draw [photon] (zM) -- (zB);
  \draw [fill] (xA) circle (.08);
  \draw [fill] (wA) circle (.08);
  \draw [fill] (xB) circle (.08);
  \draw [fill] (yB) circle (.08);
  \draw [fill] (zB) circle (.08);
  \draw [fill] (wB) circle (.08);  
  \draw [fill] (xM) circle (.08);
  \draw [fill] (zM) circle (.08);  
\end{tikzpicture}


\begin{tikzpicture}[baseline={([yshift=-1ex]current bounding box.south)},scale=.7]
 \coordinate (inA) at (0.4,.7);
  \coordinate (outA) at (3.4,.7);
  \coordinate (inB) at (0.4,-.7);
  \coordinate (outB) at (3.4,-.7);
  \coordinate (xA) at (1-0.1,.7);
  \coordinate (xxA) at (1.3,.7);
  \coordinate (yA) at (1.6,.7);
  \coordinate (zA) at (2.2,.7);
  \coordinate (wA) at (2.8+0.1,.7);
  \coordinate (xB) at (1-0.1,-.7);
  \coordinate (xxB) at (1.3,-.7);
  \coordinate (yB) at (1.6,-.7);
  \coordinate (zB) at (2.2,-.7);
  \coordinate (wB) at (2.8+0.1,-.7);
  \coordinate (xM) at (1-0.1,0);
  \coordinate (xxM) at (1.3,0);
  \coordinate (yM) at (1.6,0);
  \coordinate (yyM) at (1.9,0);
  \coordinate (zM) at (2.2,0);
  \coordinate (wM) at (2.8+0.1,0);
  \draw [dotted] (inA) -- (outA);
  \draw [dotted] (inB) -- (outB) node [midway, below] {(K3.2)};;
  \draw [photon] (xA) -- (xB);
  \draw [photon] (xA) -- (yyM);
  \draw [photon] (yyM) -- (yB);
  \draw [photon] (yyM) -- (zB);
  \draw [photon] (wA) -- (wM) -- (wB);
  \draw [photon] (yyM) -- (zB);
  \draw [photon] (yyM) -- (wM);
  \draw [fill] (xA) circle (.08);
  \draw [fill] (wA) circle (.08);
  \draw [fill] (xB) circle (.08);
  \draw [fill] (yB) circle (.08);
  \draw [fill] (zB) circle (.08);
  \draw [fill] (wB) circle (.08);  
  \draw [fill] (yyM) circle (.08);
  \draw [fill] (wM) circle (.08);  
  \end{tikzpicture}
 

\begin{tikzpicture}[baseline={([yshift=-1ex]current bounding box.south)},scale=.7]
 \coordinate (inA) at (0.4,.7);
  \coordinate (outA) at (3.4,.7);
  \coordinate (inB) at (0.4,-.7);
  \coordinate (outB) at (3.4,-.7);
  \coordinate (xA) at (1-0.1,.7);
  \coordinate (xxA) at (1.3,.7);
  \coordinate (yA) at (1.6,.7);
  \coordinate (zA) at (2.2,.7);
  \coordinate (wA) at (2.8+0.1,.7);
  \coordinate (xB) at (1-0.1,-.7);
  \coordinate (xxB) at (1.3,-.7);
  \coordinate (yB) at (1.6,-.7);
  \coordinate (zB) at (2.2,-.7);
  \coordinate (wB) at (2.8+0.1,-.7);
  \coordinate (xM) at (1-0.1,0);
  \coordinate (xxM) at (1.3,0);
  \coordinate (yM) at (1.6,0);
  \coordinate (yyM) at (1.9,0);
  \coordinate (zM) at (2.2,0);
  \coordinate (wM) at (2.8+0.1,0);
  \draw [dotted] (inA) -- (outA);
  \draw [dotted] (inB) -- (outB) node [midway, below] {(K3.3)};;
  \draw [photon] (xA) -- (xM) ;
  \draw [photon] (xB) -- (xM) ;
  \draw [photon] (xM) -- (yB);
  \draw [photon] (xM) -- (zA);
  \draw [photon] (wA) -- (wB);
  \draw [photon] (zA) -- (zB);
  \draw [zUndirected] (zA) -- (wA);
  \draw [fill] (xA) circle (.08);
  \draw [fill] (wA) circle (.08);
  \draw [fill] (zA) circle (.08);
  \draw [fill] (xB) circle (.08);
  \draw [fill] (yB) circle (.08);
  \draw [fill] (zB) circle (.08);
  \draw [fill] (wB) circle (.08);  
  \draw [fill] (xM) circle (.08);  
\end{tikzpicture}


\begin{tikzpicture}[baseline={([yshift=-1ex]current bounding box.south)},scale=.7]
 \coordinate (inA) at (0.4,.7);
  \coordinate (outA) at (3.4,.7);
  \coordinate (inB) at (0.4,-.7);
  \coordinate (outB) at (3.4,-.7);
  \coordinate (xA) at (1-0.1,.7);
  \coordinate (xxA) at (1.3,.7);
  \coordinate (yA) at (1.6,.7);
  \coordinate (zA) at (2.2,.7);
  \coordinate (wA) at (2.8+0.1,.7);
  \coordinate (xB) at (1-0.1,-.7);
  \coordinate (xxB) at (1.3,-.7);
  \coordinate (yB) at (1.6,-.7);
  \coordinate (zB) at (2.2,-.7);
  \coordinate (wB) at (2.8+0.1,-.7);
  \coordinate (xM) at (1-0.1,0);
  \coordinate (xxM) at (1.3,0);
  \coordinate (yM) at (2.2,0);
  \coordinate (yyM) at (1.9,0);
  \coordinate (zM) at (2.2,0);
  \coordinate (wM) at (2.8+0.1,0);
  \draw [dotted] (inA) -- (outA);
  \draw [dotted] (inB) -- (outB) node [midway, below] {(K3.4)};;
  \draw [photon] (xA) -- (xM) -- (xB);
  \draw [photon] (xA) -- (yM);
  \draw [photon] (yM) -- (yB);
  \draw [photon] (wA) -- (wB);
  \draw [photon] (zA) -- (zB);
  \draw [zUndirected] (zA) -- (wA);
  \draw [fill] (xA) circle (.08);
  \draw [fill] (wA) circle (.08);
  \draw [fill] (zA) circle (.08);
  \draw [fill] (xB) circle (.08);
  \draw [fill] (yB) circle (.08);
  \draw [fill] (zB) circle (.08);
  \draw [fill] (wB) circle (.08);  
  \draw [fill] (yM) circle (.08);  
\end{tikzpicture}
  

\begin{tikzpicture}[baseline={([yshift=-1ex]current bounding box.south)},scale=.7]
 \coordinate (inA) at (0.4,.7);
  \coordinate (outA) at (3.4,.7);
  \coordinate (inB) at (0.4,-.7);
  \coordinate (outB) at (3.4,-.7);
  \coordinate (xA) at (1,.7);
  \coordinate (xxA) at (1.3,.7);
  \coordinate (yA) at (1.6,.7);
  \coordinate (zA) at (2.2,.7);
  \coordinate (wA) at (2.8,.7);
  \coordinate (xB) at (1,-.7);
  \coordinate (xxB) at (1.3,-.7);
  \coordinate (yB) at (1.6,-.7);
  \coordinate (zB) at (2.2,-.7);
  \coordinate (wB) at (2.8,-.7);
  \coordinate (xM) at (1,0);
  \coordinate (xxM) at (1.3,0);
  \coordinate (yM) at (1.6,0);
  \coordinate (yyM) at (1.9,0);
  \coordinate (zM) at (2.2,0);
  \coordinate (zzM) at (2.5,0);
  \coordinate (wM) at (2.8,0);
  \draw [dotted] (inA) -- (outA);
  \draw [dotted] (inB) -- (outB) node [midway, below] {(K3.5)};;
  \draw [photon] (xA) -- (xxM) -- (xB);
  \draw [photon] (wA) -- (zzM) -- (wB);
  \draw [photon] (xxM) -- (zzM);
  \draw [photon] (xxM) -- (yB);
  \draw [photon] (zzM) -- (zB);
  \draw [fill] (xA) circle (.08);
  \draw [fill] (wA) circle (.08);
  \draw [fill] (xB) circle (.08);
  \draw [fill] (yB) circle (.08);
  \draw [fill] (zB) circle (.08);
  \draw [fill] (wB) circle (.08);  
  \draw [fill] (xxM) circle (.08);
  \draw [fill] (zzM) circle (.08);  
\end{tikzpicture}

}

\raisebox{-0.55cm}{


\quad \,\,
 \raisebox{-0.3cm}{
\begin{tikzpicture}[baseline={([yshift=-1ex]current bounding box.south)},scale=.7]
  \coordinate (inA) at (0.4,.7);
  \coordinate (outA) at (4.35,.7);
  \coordinate (inB) at (0.4,-.7);
  \coordinate (outB) at (4.35,-.7);
  \coordinate (xA) at (1,.7);
  \coordinate (xxA) at (1.825,.7) ;
  \coordinate (xxB) at (1.825,-.7) ;
  \coordinate (yA) at (1.60,.7);
  \coordinate (yyA) at (2.375,0) ;
  \coordinate (zA) at (2.925,.7);
  \coordinate (zzA) at (3.75,.7) ;
  \coordinate (xB) at (1-0.2,-.7);
  \coordinate (yB) at (1.5,-.7);
  \coordinate (zB) at (2.925,-.7);
  \coordinate (zzB) at (3.75+0.2,-.7);
  \coordinate (xM) at (1,0);
  \coordinate (yM) at (1.5,0);
  \coordinate (zM) at (3,-0);
  \draw [dotted] (inA) -- (outA);
  \draw [dotted] (inB) -- (outB) node [midway, below] {(CY3)};
  \draw [draw=none] (xA) to[out=40,in=140] (zA);
  \draw [photon] (xxA) -- (xB);
  \draw [photon] (yyA) -- (zB);
  \draw [photon] (zA) -- (zzB);
  \draw [photon] (yyA) -- (xxB);
  \draw [photon] (xxA) -- (yyA);
   \draw [photon] (zA) -- (yyA);
  \draw [fill] (xxB) circle (.08);

  \draw [fill] (yyA) circle (.08);
  \draw [fill] (xxA) circle (.08);
  \draw [fill] (zA) circle (.08);
  \draw [fill] (zB) circle (.08);
  \draw [fill] (xB) circle (.08);
  \draw [fill] (zzB) circle (.08);
\end{tikzpicture}
}}
5PM-2SF:
\raisebox{-0.9cm}{

\begin{tikzpicture}[baseline={([yshift=-1ex]current bounding box.south)},scale=.7]
  \coordinate (inA) at (0.4,.7);
  \coordinate (outA) at (3.6,.7);
  \coordinate (inB) at (0.4,-.7);
  \coordinate (outB) at (3.6,-.7);
  \coordinate (xA) at (1,.7);
  \coordinate (xxA) at (1.825,.7) ;
  \coordinate (xxB) at (1.825,-.7) ;
  \coordinate (yA) at (1.60+0.4,.7);
  \coordinate (yyA) at (2.375,0) ;
  \coordinate (zA) at (3,.7);
  \coordinate (zzA) at (3.75,.7) ;
  \coordinate (xB) at (1,-.7);
  \coordinate (yB) at (1.6+0.4,-.7);
  \coordinate (zB) at (3,-.7);
  \coordinate (zzB) at (3.75,-.7);
  \coordinate (xM) at (1,0);
  \coordinate (yM) at (1.6+0.4,0);
  \coordinate (zM) at (3,-0);
  \draw [dotted] (inA) -- (outA);
  \draw [dotted] (inB) -- (outB) node [midway, below] {(K3')};
  \draw [draw=none] (xA) to[out=40,in=140] (zA);
  \draw [photon] (xA) -- (xB);
  \draw [photon] (yA) -- (yM) -- (yB);
  \draw [photon] (zA) -- (zB);
  \draw [photon] (xA) -- (yM) -- (zB);
  \draw [fill] (yM) circle (.08);

  \draw [fill] (xA) circle (.08);
  \draw [fill] (yA) circle (.08);
  \draw [fill] (zA) circle (.08);
  \draw [fill] (xB) circle (.08);
  \draw [fill] (yB) circle (.08);
  \draw [fill] (zB) circle (.08);
\end{tikzpicture}

\quad \,\,

\begin{tikzpicture}[baseline={([yshift=-1ex]current bounding box.south)},scale=.7]
  \coordinate (inA) at (0.4,.7);
  \coordinate (outA) at (3.6,.7);
  \coordinate (inB) at (0.4,-.7);
  \coordinate (outB) at (3.6,-.7);
  \coordinate (xA) at (1,.7);
  \coordinate (xxA) at (1.825,.7) ;
  \coordinate (xxB) at (1.825,-.7) ;
  \coordinate (yA) at (1.60+0.4,.7);
  \coordinate (yyA) at (2.375,0) ;
  \coordinate (zA) at (3,.7);
  \coordinate (zzA) at (3.75,.7) ;
  \coordinate (xB) at (1,-.7);
  \coordinate (yB) at (1.6+0.4,-.7);
  \coordinate (zB) at (3,-.7);
  \coordinate (zzB) at (3.75,-.7);
  \coordinate (xM) at (1,0);
  \coordinate (yM) at (1.6+0.4,0);
  \coordinate (zM) at (3,-0);
  \draw [dotted] (inA) -- (outA);
  \draw [dotted] (inB) -- (outB) node [midway, below] {(CY3')};
  \draw [draw=none] (xA) to[out=40,in=140] (zA);
  \draw [photon] (xA) -- (xB);
  \draw [photon] (yA) -- (xM);
  \draw [photon] (zA) -- (zB);
  \draw [photon] (xM) -- (yM) -- (zM);
  \draw [photon] (yB) -- (zM);
  \draw [fill] (xM) circle (.08);
   \draw [fill] (zM) circle (.08);

  \draw [fill] (xA) circle (.08);
  \draw [fill] (yA) circle (.08);
  \draw [fill] (zA) circle (.08);
  \draw [fill] (xB) circle (.08);
  \draw [fill] (yB) circle (.08);
  \draw [fill] (zB) circle (.08);
\end{tikzpicture}

\qquad

}

\caption{\small The graphs up to 5PM whose sectors are associated to CY manifolds. 
Here, only either the even or odd parity integrals give rise to the CY period integrals. The top
line corresponds to the same K3 surface, cf.~table \ref{K3Ints}. Dotted lines represent $\delta$-functions, solid lines the linear propagators $D_1,\hdots, D_4$ and wiggly lines the graviton propagators $D_5,\hdots,D_{18}$.}
  \label{fig2}
\end{figure*}

\section{Integral family}
The generation of the WQFT integrand has been described at 3PM and 4PM orders in \cite{Jakobsen:2022fcj,Jakobsen:2022psy,Jakobsen:2022zsx,Jakobsen:2023oow,Jakobsen:2023ndj,Jakobsen:2023hig}. 
It employs recursive diagrammatic techniques and tensor reduction for the generation of the
WQFT integrands. In the 4PM case the emerging integral family is comprised of 12 propagators
and three worldline delta functions  \cite{Jakobsen:2023ndj}. 
At the  5PM-1SF order, this automated integrand generation is, in principle, identical
and poses no technical problems \cite{Driesse:2024xad} from which
the scalar integral families may be read off.
\begin{widetext}
We encounter an integral family comprised of 
18 propagators and four worldline delta functions taking the form
\begin{subequations}
  \label{eq:general}
  \begin{align}
    I^{(\sigma_1,\sigma_2,...,\sigma_7)}_{n_1,n_2,...,n_{22}}
    =
    \int_{\ell_1,\ell_2,\ell_3,\ell_4}\!\!\!\!\!\!\!\!\!\!\!\!\!\!
    \frac{
      \dd^{(n_1-1)}(\ell_1\cdot v_{2})
      \dd^{(n_2-1)}(\ell_2\cdot v_{2})
      \dd^{(n_3-1)}(\ell_3\cdot v_{2})
       \dd^{(n_4-1)}(\ell_4\cdot v_{1})
    }{
      D_1^{n_5}
      D_2^{n_6}
      \cdots
      D_{18}^{n_{22}}
    } \, ,
  \end{align}
  where $\delta^{(n)}(x)$ denotes the $n$th derivative of the worldline delta function.
 The propagators ($k=1,2,3$ and $j=1,2,3,4$) are given by
 \begin{align}
    &
    D_k
    =
    \ell_{k}\cdot v_{1} +\sigma_{k}i0^+
    \,,\,
    D_4
    =
    \ell_{4}\cdot v_{2} +\sigma_{4}i0^+\,,\,
    D_{4+k}
    =
    (\ell_k-\ell_4)^2+\sigma_{4+k}\,\text{sgn}\,(\ell_k^0-\ell_4^0)\,i0^+
    \,,\,\,\\
    &D_{8}
    =
    (\ell_1-\ell_2)^2
    \, , \,
    D_{9}
    =
    (\ell_1-\ell_3)^2
    \, , \,
    D_{10}
    =
    (\ell_2-\ell_3)^2
    \, , \,
   D_{10+j}
   =\ell_j^2
   \,,\,
   D_{14+j}
   =(\ell_j+q)^2
   \,.\nn
  \end{align}
\end{subequations}
In principle, all graviton propagators carry a retarded $i0$ prescription. Due to the delta functions only three propagators can go on-shell. The sign of the $i0$ prescription is defined by $\sigma_i= \pm 1$.
The 5PM-1SF integral family splits into two branches: even ($b$-type) or odd ($v$-type) in parity ($v \rightarrow -v$), determined by the number of worldline propagators and derivatives of the delta functions, i.e.~the
parity of the first eight indices $\{n_1,\ldots n_8\}$. 
These two integral branches couple in the final result to the vectors $b^\mu$ and $v_i^\mu$ respectively. The integrals of $b$- and $v$-type can have very different function spaces. 
Notice that all $v$-type integrals without a worldline propagator vanish, when using Feynman propagators, due to symmetries ($l_i\rightarrow -l_i, q\rightarrow -q$). Therefore, they only contribute to the dissipative part of the impulse.
\end{widetext}
The usage of retarded propagators makes the even integrals purely real and the odd integrals pseudo-real. The integrals are effectively one-scale integrals depending on the parameter 
$x\in[0,1]$. The self-force order determines the indices of the velocities in the delta functions. At $n$PM-$m$SF order, we have $(n-1-m)$-loop momenta contracted with $v_1$ and $m$ momenta contracted with $v_2$ in the delta functions. 
At 0SF order, the $\gamma$ dependence becomes trivial. The complexity of the integration problem increases with every self-force order.

\section{IBP reduction and choice of basis}
To study the function space of the PM integrals one necessary step is to derive the differential equations of the involved MIs. The MIs as well as their differential equations are derived from IBP relations using the program {\tt Kira} \cite{Maierhofer:2017gsa,Klappert:2020nbg,Lange:2021edb}. To simplify this task we only reduced the derivatives of the MIs neglecting the full integrand reduction.

It is convenient to group the MIs into a large vector $\underline{I}(x,\epsilon)$. Here, we order MIs from lower to higher sectors, i.e.~we start with the sub-sectors. With this convention we find that the GM equation takes the form $(\mathrm d - M(x,\epsilon))\underline{I}(x,\epsilon)=0,$ 
such that the connection matrix $M(x,\epsilon)$ factorizes into sectors and is of lower block triangular form. As discussed before, we split the integrals into even and odd parity and look at them separately. The diagonal blocks correspond to the maximal cuts of the system~\cite{Primo:2016ebd,Primo:2017ipr}. They essentially determine the class of transcendental functions appearing in a given sector. 

To systematically solve the GM equation up to a given order in $\epsilon$ it is useful to go to an $\epsilon$-factorized differential equation \cite{Henn:2013pwa}. For this one has to construct a rotation into a new basis $\underline{J}(x,\epsilon)=T(x,\epsilon)\underline{I}(x,\epsilon)$ such that the $\epsilon$ dependence is factored out in the new connection matrix
\begin{equation}
\begin{aligned}
    0 &= (\mathrm d - \epsilon A(x))\underline{J}(x,\epsilon)  \, , \\
    \epsilon A(x) &= (T(x,\epsilon) M(x,\epsilon) + \mathrm dT(x,\epsilon))T(x,\epsilon)^{-1} \, .
\label{epsgm}
\end{aligned}
\end{equation}
The complexity of calculating the rotation $T(x,\epsilon)$ strongly depends on the initial choice of MIs. Essential criteria for a good selection of MIs are the absence of power-like singularities, the absence of general polynomials $p(x,\epsilon)$ in the denominators of $M(x,\epsilon)$ and the manifestation of the different appearing minimally coupled systems, i.e.~geometries, in the problem. For our case, this means that the choice of MIs gives rise either to projective spaces with marked points (polylogarithms), CY manifolds or possible additional residues on these geometries~\cite{Gorges:2023zgv}. For GM systems related to CY $n$-folds, the main ingredient then is to use GT to construct linear combinations of MIs involving the CY period integrals as coefficients such that their leading singularities satisfy unipotent differential equations~\cite{Neganew}. Practically, this means that one has to split the Wronskian matrix of fundamental solutions $W(x)$ into a semi-simple and unipotent part, i.e. $W(x) = W(x)^\text{ss} W(x)^\text{u}$. After removing the semi-simple part from the initial MIs and further total derivatives one arrives at an $\epsilon$-factorized differential equation. In some cases, additional new transcendental functions being iterated integrals of the CY period integrals have to be introduced during these steps. For more details, we refer to~\cite{Gorges:2023zgv,Neganew}.

The final $\epsilon$-factorized connection matrix $A(x)$ consists of rational functions and iterated CY period integrals. We can see now clearly that the higher $\epsilon$-orders just give further iterated integrals of these kernels. Also the contributions from sub-sectors do not change the function space.

\section{CY in the Sky}
The rank $n+1$ GM connection $(\mathrm d-A^M(x))\underline \Pi(x) = 0$  of a one-parameter period motive of a CY $n$-fold $M$ appears 
as sub-connection in \eqref{epsgm}, i.e. as a block in $M(x,0)$. It can be equivalently written as a linear Picard-Fuchs differential operator (PF op) of order $n+1$, i.e. ${\mathcal L}^{(n+1)}=\partial_x^{n+1}+ \sum_{i=0}^{n} a_i(x) \,\partial_x^i$ where 
$a_i(x)$ is a rational function in $x$. The solutions ${\underline \Pi}(x)$ to ${\mathcal L}^{(n+1)}{\underline \Pi}(x)=0$  are then the periods $\Pi_k=\int_{C^k_n} \Omega$  of $M$, with $\Omega$ the holomorphic 
$(n,0)$-form and the $n$-cycles  $C^k_n$ can be chosen to be in $H_n(M,\mathbb{Z})$.

In order to describe a period motive associated to 
an $n$-dimensional algebraic variety, ${\mathcal L}^{(n+1)}$ has to have only regular singular points of maximal unipotency $n$ in its moduli space ${\mathcal M}_x$. For ${\mathcal L}^{(n+1)}$ to be also a CY operator, GT requires that ${\mathcal L}^{(n+1)}$ has to be self-adjoint, i.e.
\be  
{\mathcal L}^{* (n+1)}c(x)=(-1)^{n+1}c(x){\mathcal L}^{(n+1)}\, .
\label{eq:selfadjointness}
\ee   
Here, ${\mathcal L}^{* (n+1)}=\sum_{i=0}^{n+1}\left(-\partial_x \right)^i a_i(x)$ is the adjoint operator
and the rational function $c(x)$ is determined by $\partial_x c(x)/c(x)=2 a_n(x)/(n+1)$ up 
to a multiplicative  constant. CY motives are defined by CY differential operators, see~\cite{MR3822913} for $n=3$, which have in addition a point of maximal unipotent monodromy (MUM point). At this point they have an integral 
mirror map and integral BPS expansions \cite{Aspinwall:1991ce,MR1416356,Klemm:2007in}. The latter are trivial for K3 surfaces, and in this case, the period domain is a symmetric domain. Moreover, the periods are modular 
functions of congruent subgroups of  SL$(2,\mathbb{Z})$, which is related to the fact that one-parameter K3 period integrals are symmetric squares of elliptic integrals~\cite{Lerche:1991wm,MR1411339,Klemm:2019dbm,Broedel:2019kmn}. For integral 
BPS expansion in three- and four-folds see \cite{Aspinwall:1991ce,Klemm:1996ts,Mayr:1996sh,Klemm:2007in}.
The integral stuctures are consequences of the integral monodromy 
respresentation of $O(\Sigma,\mathbb{Z})$ for $n$ even and 
$\mathrm{Sp}(2n+2,\mathbb{Z})$ for $n$ odd which also encode the topological type of $M$. Here $\Sigma$ is the intersection form on $H_n(M,\mathbb{Z})$.

To identify a given graph $\Gamma$ in the 5PM expansion having $m$ different MIs in its sector with a CY $n$-fold, it is essential that the corresponding MIs are chosen such that at $\epsilon=0$ the candidate minimally coupled CY block is decoupled from other contributions which can be additional residues, polylogarithmic contaminations or additional non-trivial CY manifolds~\cite{Gorges:2023zgv,Neganew}. This means that at $\epsilon=0$ the connection form of this sector splits as 
\begin{equation}
\begin{aligned}
    M_\Gamma(x,0) =  \begin{pmatrix} A^{M_\Gamma}(x)&  0 \\ C(x) & D(x) \end{pmatrix}\, ,
\label{CYref}
\end{aligned}
\end{equation}
where the two matrices $C(x), D(x)$ describe the additional structures in the sector. Then, the CY can be identified directly from $A^{M_\Gamma}$ or via the equivalent PF op. Alternatively, a  Baikov representation associated to 
$\Gamma$ can be expanded in $x$  by performing a suitable 
residuum calculation to high order, which corresponds to choose $C_n=T^n$ to obtain a
torus integral $\int_{T^n} \Omega$  in $M_\Gamma$. An ansatz 
for ${\mathcal L}^{(n+1)}(x)$ can, in practice, be uniquely fixed  by the requirement 
 ${\mathcal L}^{(n+1)}(x) \int_{T^n} \Omega=0$ and can be compared with \eqref{CYref}.

In all cases, we find the higher rank period motives of $M_\Gamma$   fulfilling 
 the conditions above  to be  CY motives.  They determine 
the corresponding higher transcendental functions, encoding the observables of the black hole scattering process, which are 
CY periods $\underline \Pi(x)$ and their extensions as we will explicitly exemplify below.


\section{The 5PM-1SF sector}
We have found five different graphs (K3.$i$, $i=1,\hdots,5$ in fig \ref{fig2})  related to a K3 surface. To make this identification, a suitable choice of MIs is given by the corner integrals listed in table \ref{K3Ints}. Here, each corner integral corresponds to a single graph  and is supplemented by additional MIs being the derivatives of the corner one. In this regard, the graphs K3.1-2 consist of four MIs whereas the graphs K3.3-5 consist of three MIs. 
  
\begin{table}[h]
\begin{tabular}{c|c|c}
MI & $n_{i}$ & parity\\
\hline
$I_{\mathrm{K3},1}$ & 1 1 1 1 0 0 0 -1 0 1 1 1 0 1 1 0 0 1 0 0 1 0 & odd\\
$I_{\mathrm{K3},2}$ & 1 1 1 1 0 0 0 -1 1 1 0 1 0 1 1 0 0 0 0 0 1 1 & odd\\ 
$I_{\mathrm{K3},3}$ & 1 1 1 1 0 0 1  0 0 1 0 1 0 1 1 0 0 1 0 0 1 0& odd\\
$I_{\mathrm{K3},4}$ & 1 1 1 1 0 0 1  0 1 0 1 1 0 1 1 0 0 0 0 0 1 0& odd\\
$I_{\mathrm{K3},5}$ & 1 1 1 1 0 0 0  0 0 1 0 1 0 1 1 0 0 1 0 0 1 1& even\\
\end{tabular}
\caption{The master integrals of the K3 sectors.}
\label{K3Ints}
\end{table}
  
For these choices of MIs, the corresponding GM systems can be determined from IBP relations. It is interesting to observe, that all K3 operators appearing in all five graphs at $\epsilon=0$ are related to the same K3 operator also appearing in the 4PM-1SF sector (graph K3.0). This operator is conveniently expressed as
\be
{\mathcal L}^{(3)}_1 = (2\theta -1)^3+z^2 (2\theta +1)^3- 4 z\theta(4 \theta^2+1)    
\label{eq:op0} 
\ee 
with $\theta=z \frac{\rm d}{{\rm d} z}$, $z=x^2$. Self-adjointness of ${\mathcal L}^{(3)}_1$ is guaranteed through $c(z)= 4/(z (1-z)^2)$ from \eqref{eq:selfadjointness}, 
where the $4$ has been determined from the intersection $\Sigma$ 
of the K3. Monodromy properties can be read off from the corresponding  Riemann $\mathcal P$-Symbol in the appendix eq.~\eqref{eq:riemannsymbolK3}. 
GT and the representation theory of the monodromy groups~\cite{Lerche:1991wm,MR1411339,Klemm:2019dbm} imply that ${\mathcal L}^{(3)}={\rm Sym}^{2} ({\mathcal L}^{(2)})$, where $\mathcal{L}^{(2)}$ 
is the PF op of an elliptic curve. In our case, it is the Legendre curve $Y=X(X-1)(X-z)$ with monodromy group $\Gamma_0(4)$ and ${\mathcal L}^{(2)}_1= \theta^2   - z (\theta+\h)^2$. The K3 geometry is then the twisted product of the latter given by 
\be 
Y^2=  X(X-1) (X-z) Z (Z-1)(Z-1/z)\, .
\label{eq:K3one}
\ee 
Its symmetry makes it immediately clear that the same solution 
structure appears at $w=1/z=1/x^2$. This symmetry is 
inherited from the physical parametrization $\gamma=(x+x^{-1})/2$ 
and must occur in all geometries. Since elliptic curves cannot exhibit this symmetry in their moduli space ${\mathcal M}_z$ the occurrence of CY motives in the PM approximation starts with $n=2$, i.e. K3 surfaces. 
We can bring all sectors corresponding to the five graphs K3.1-5 in fig.~\ref{fig2} into $\epsilon$-form using the \texttt{INITIAL} algorithm \cite{Dlapa:2022wdu,Dlapa:2020cwj}. The corner integrals in table \ref{K3Ints} serve up to normalization as initial integrals for the \texttt{INITIAL} algorithm.

\begin{table}[h]
\begin{tabular}{c|c|c}
MI & $n_{i}$  & parity\\
\hline
$I_{1}$ & 1 1 1 2 0 0 0 0 1 0 1 1 0 1 1 0 0 0 0 0 1 0 & odd  \\
$I_{5}$ & 1 1 2 1 0 0 0 0 1 0 1 1 0 1 1 0 0 0 0 0 1 0 & odd \\ 
$I_{6}$ & 1 1 2 2 0 0 0 -1 1 0 1 1 0 1 1 0 0 0 0 0 1 0& odd \\
\end{tabular}
\caption{The master integrals of the CY3 sector.}
\label{table1}
\end{table}

Besides the K3 surface there is only one other CY $n$-fold appearing in the 5PM-1SF sector. The graph inducing this (first) CY three-fold is depicted in fig.~\ref{fig2} (CY3).
While its even sub-sector is polylogarithmic (as reported in \cite{RufAmps}) the odd sub-sector is not and will contribute to the \emph{dissipative} part of the 5PM-1SF impulse. This CY three-fold sector is built up of six MIs. The first four MIs, which are the corner integral $I_1$ in table \ref{table1} and its three derivatives, describe the CY three-fold part. The MIs $I_5,I_6$ are instead additional residues on that CY. The corresponding fourth-order CY operator $\mathcal L^{(4)}_1$ is of hypergeometric type and is given by
\be
    \mathcal L^{(4)}_1 = \theta^4 - 2^8 z (\theta + \h)^4
\label{eq:PF2222}   
\ee
in the variable $z = 2^{-8} x^4$ and after normalizing $I_1$ by $x$ with $c(z)=16/(z^3 (1-2^8 z))$ from \eqref{eq:selfadjointness}. ${\cal L}^{(4)}_1$ is a Hadamard product ${\cal L}^{(2)}_1 * {\cal L}^{(2)}_1$ of the Legendre operator, see \cite{MR4668024} for Hadamard constructions. 
The  corresponding smooth CY three-fold one-parameter complex family  
$z= (2 \psi)^{-8}$, can be defined as resolution of four symmetric quadrics 
\be 
x_j^2+y_j^2-2 \psi x_{j+1} y_{j+1}=0,\ j\in \mathbb{Z}/4 \mathbb{Z}
\label{eq:CY1}
\ee
in the homogeneous coordinates $x_i,y_j$, $j=0,\ldots,3$  of $\mathbb{P}^7$~\cite{Bonisch:2022mgw}. This hypergeometric CY 
three-fold motive appeared in the study of mirror symmetry in~\cite{MR1201748}.

To derive the $\epsilon$-factorization of this CY block, we first have to split the matrix of fundamental solutions $W(x) = (\partial^j\varpi_i)_{0\leq i,j\leq 3}$ into its semi-simple and unipotent part (for more details see the appendix). The unipotent part satisfies
\begin{equation}
    (\mathrm d-A^\text{u}(x))W^\text{u}(x)=0 \, , \quad  A^\text{u}(x) = \begin{pmatrix} 0 & 1 & 0 & 0 \\ 0 & 0 & Y_1 & 0 \\ 0 & 0 & 0 & 1 \\
0 & 0 & 0 & 0 \end{pmatrix}
\end{equation}
and $A^\text{u}(x)$ is nilpotent, i.e. $(A^\text{u}(x))^{4}=0$. From the CY perspective it is meaningful to use the rescaled variable $\mathrm x=\frac x4$ in which the integral expansions are manifest. 
After removing the semi-simple part from the initial MIs, we have to introduce four new transcendental functions, which are iterated integrals of CY periods, to obtain the $\epsilon$-form. The two simplest ones are given by
\begin{equation}
\begin{aligned}
 G_1(\mathrm x) &= -\int_0^\mathrm x \frac{24576 \mathrm x' \left(1+256 \mathrm x'^4\right)}{\left(1-256 \mathrm x'^4\right)^2} \frac{\varpi_0(\mathrm x')^2}{\alpha_1(\mathrm x')}\mathrm d\mathrm x' \, , \\
 G_3(\mathrm x) &= \int_0^\mathrm x \frac{\mathrm x'}{1-256 \mathrm x'^4} \frac{G_1(\mathrm x')\alpha_1(\mathrm x')^2}{\varpi_0(\mathrm x')^2}\mathrm d\mathrm x' \, .
\end{aligned}
\end{equation}
The $\mathrm x$ expansions of $\alpha_1$ and all $G_i$ functions are given in the appendix in eq. \eqref{eq:alpha1} and \eqref{Gexpansions}, respectively. This allows to construct the full $\epsilon$-factorized connection matrix in the CY sector by using the special properties of the CY geometry, in particular, GT and the period integrals. The new transcendental functions are given as power series which can be easily analytically continued to the whole complex plane. As an important observation, the new transcendental functions have all integer coefficient expansions in $\mathrm x$, similarly to the novel transcendental functions appearing in
~\cite{Gorges:2023zgv,Neganew,Pogel:2022yat,Pogel:2022ken,Pogel:2022vat} which for K3 surfaces are related to magnetic modular forms~\cite{Broadhurst:2017lzf,Pasol:2020twk,Neganew}. This gives us full analytic control over the function space in the 5PM-1SF sector \cite{us}. In the appendix, we provide the $\epsilon$-factorized differential equation for the CY three-fold in expanded form (see eqs. \eqref{epsformCY3} and \eqref{kernelsCY3}) together with a plot (see fig. \ref{fig:plots}) showing explicitly the higher $\epsilon$-orders.

\section{Outlook to 5PM-2SF}
This sector contains many more and new CY $n$-folds compared to the 1SF sector. For instance, for the graph K3' in fig. \ref{fig2} we find among the nine MIs a K3 
block that leads with $z=x^2$, removing an apparent singularity at $z=1$ and normalizing the corresponding MIs by $z/(1-z)$ to the famous Ap\`ery operator 
\be 
\mathcal{L}^{(3)}_A = \theta ^3+z^2 (\theta +1)^3 -z (2 \theta +1) \left(17 \theta ^2+17 \theta +5\right) 
\label{eq:apery} 
\ee
with $c(z)=\frac{\kappa}{z^2 \left(z^2-34 z+1\right)}$ from \eqref{eq:selfadjointness}, that was 
used in~\cite{MR3363457} to prove the irrationality of 
$\zeta(3)$, see \cite{MR3890449} for a review. The smooth K3 of Picard rank 19 was described in~\cite{MR0749676} as the resolution of the affine equation
\be 
1-(1- X Y )Z - z X Y Z (1-X)(1- Y)(1-Z)=0 \ . 
\label{eq:BPK3}   
\ee  
In fact, the corresponding Baikov integral representation 
and the evaluation of $\varpi_0$ near $z=0$ by performing 
the $T^2$ integral can be also found in~\cite{MR0749676}.

In recent work \cite{Frellesvig:2023bbf} a singular Baikov integral representation 
for the graph CY3' (see fig. \ref{fig2}) was given. It can be viewed
as a double cover of $\mathbb{P}^3$ branched at 
\begin{equation}
\begin{aligned}  
P =&t^2(W X+Y^2)^2(W+Z)^2(X+Z)^2 \\[2 mm] & +  2^6(1+t)(WXY)^2 Z(W+X+Z) 
\end{aligned}
\label{eq:P}
\end{equation}
with $t=x^2-1$. The corresponding geometry $U^2=P$ 
is highly singular and was not 
resolved in~\cite{Frellesvig:2023bbf} to a CY unlike \eqref{eq:BPK3} in~\cite{MR0749676}. Also the differential operator given in~\cite{Frellesvig:2023bbf} with an apparent singularity, is not of CY type~\cite{MR3822913}. It lacks integral BPS expansions at the MUM point $t=0$. With $z=-\frac{t^2}{2^{12}(1+t)}$, we lift the apparent singularity and in the appendix eq.~\eqref{eq:riemannsymbolothers} we show that 
the transformed operator
\begin{equation}
\begin{aligned}  
\mathcal{L}_2^{(4)}\! =\! \theta^4\!-\!2^{30} z^3(\theta+\h)^4 \!-\! 2^4 z(192 \theta^4\!+\! 128\theta^3\!+\!112 \theta^2 \\
+48 \theta+7)\!+\!2^{14} z^2 (192 \theta^4\!+\! 256 \theta^3\!+\!208 \theta^2\!+\!64 \theta\!+\!7)
\end{aligned}
\label{eq:PF2CY3} 
\end{equation}
corresponds to a one-parameter CY family with topological data $\chi=80$, $\kappa=4$ and $\gamma=\kappa (6 m-5)$, which fixes the topological type 
according to \cite{MR0215313}. Using integral BPS expansion we can also relate it to a Hadamard construction.  

\section{Conclusions}
In this work, we have shown that the function space describing the radiated momentum of a scattering encounter of two BHs involves CY three-folds starting at the 5PM order.  
We have completely resolved the non-polylogarithmic function space at the 5PM-1SF order by 
$\epsilon$-factorizing the differential equations for the MIs in these sectors. In addition,
an exemplary outlook into two CY sectors at 2SF was given. We established that all CY $n$-fold periods appearing so far are either symmetric or Hadamard products of elliptic functions. Clearly, the appearance of these functions indicates that the use of advanced mathematics is necessary to address the classical two-body problem
in general relativity (in the PM or SF expansions).
While the Newtonian problem is famously linked to elliptic integrals, it is fascinating to
see that the general relativistic problem leads to their natural generalizations in terms
of CY periods.
\medskip
\pagebreak

\begin{acknowledgments}
We thank Mathias Driesse, Gustav Uhre Jakobsen, Gustav Mogull and Johann Usovitsch for ongoing collaborations as
well as Kilian B\"ohnisch, Claude Duhr, Johannes Henn, Sheldon Katz,  Nikolaos Syrrakos and Lorenzo Tancredi for insightful discussions. 
We are grateful to  Sebastian P\"ogel and Matthias Wilhelm for communications. AK also likes  to thank Ilia Gaiur, Vasily Golyshev, Minxin Huang, Matt Kerr, David Prinz, Wadim Zudilin  and other members of Group de Travail on Differential Equations for discussions and communications. 
Special thanks to  Duco van Straten for pointing out reference~\cite{MR4668024}. This work was funded by the Deutsche Forschungsgemeinschaft
(DFG, German Research Foundation)
Projektnummer 417533893/GRK2575 ``Rethinking Quantum Field Theory'' (BS)
and  by the European Union through the 
European Research Council under grant ERC Advanced Grant 101097219 (GraWFTy) (JP) and ERC Starting Grant 949279 (HighPHun) (CN).
Views and opinions expressed are however those of the authors only and do not necessarily reflect those of the European Union or European Research Council Executive Agency. Neither the European Union nor the granting authority can be held responsible for them.
\end{acknowledgments}


\appendix
\begin{widetext}
\end{widetext}

\section{Further properties of the CY operators}
In this appendix, we list more properties of the CY manifolds appearing at 5PM. We start with the Riemann $\mathcal P$-symbols of the CY operators in the 5PM-1SF sector
\be 
\mathcal P_{\mathcal L^{(3)}_1}	\begin{Bmatrix}					
						            0 & 1 & \infty  \\ \hline \\[-2ex] 
						            \frac12 & 0     & \frac12         \\[0.5ex]
						            \frac12 & 0     & \frac12         \\[0.5ex]
						            \frac12 & 0     & \frac12         \\[0.5ex]
						        \end{Bmatrix}  \, , \ %
\mathcal P_{\mathcal L^{(4)}_1}	\begin{Bmatrix}					
						            0 & \frac{1}{2^8} & \infty  \\[0.5ex] \hline \\[-2ex] 
						            0 & 0     & \frac12         \\[0.5ex]
						            0 & 1     & \frac12         \\[0.5ex]
						            0 & 1     & \frac12         \\[0.5ex]
						            0 & 2     & \frac12         \\[0.5ex]
						        \end{Bmatrix} \, ,
\label{eq:riemannsymbolK3}
\ee
and in the 2SF sector
\be 
\mathcal P_{\mathcal L^{(3)}_B}	\begin{Bmatrix}					
						            0 & p   & \frac1p & \infty  \\[1ex] \hline \\[-2ex] 
						            0 & 0   & 0       & 1       \\[0.5ex]
						            0 & \frac12   & \frac12       & 1       \\[0.5ex]
						            0 & 1   & 1       & 1       \\[0.5ex]
						        \end{Bmatrix}  \, , \ %
\mathcal P_{\mathcal L^{(4)}_2}	\begin{Bmatrix}					
						           0&  \frac{1}{2^{10}} & \infty  \\[1ex] \hline \\[-2ex] 
						            0 & 0 & \h\\[0.5ex]
						             0 & \h & \h\\[0.5ex]
						              0 & \frac{3}{2} & \h\\[0.5ex]
						               0 & 2 & \h\\[0.5ex]
						        \end{Bmatrix} \, ,
\label{eq:riemannsymbolothers}
\ee
where $p,p^{-1}$ are the two roots of the quadratic equation $u^2-34 u+1=0$. A Riemann $\mathcal P$-symbol $\mathcal P_{\mathcal L^{(r)}}$ records the local exponents $\gamma_i$, $i=1,\ldots,r$ of the solutions to a $r$th-order operator ${\mathcal L^{(r)}}(z)$, below the regular singular points or the apparent singularities in the moduli space parametrized by $z$. If all $\gamma_i$ are equal, $z$ is a MUM point. 

At a MUM point a Frobenius basis with local exponent $\gamma_1$ is given by $\underline{\tilde \Pi}=(\varpi_i,i=0,\ldots,n)$ with $\varpi_j(z)=z^{\gamma_1}\sum_{k=0}^n \frac{1}{k!} \log(z)^k f_{n-k}(z)$, where the power series are normalized like $f_0(0)=1$, $f_{i>0}(0)=0$. Then, we get, e.g. for a three-fold,  an integral symplectic basis~\cite{Hosono:1994ax,Bonisch:2020qmm,Bonisch:2022mgw} by 
\be 
{\underline \Pi}= 
\left(\begin{array}{cccc}
1& 0&0&0\\ 
0& \frac{1}{2 \pi i}&0&0\\
\frac{c_2}{24}&\frac{\sigma}{2 \pi i}&\frac{\kappa}{4\pi^2}&0\\
\frac{\chi \zeta(3)}{2(2 \pi i)^3} & \frac{c_2}{24(2 \pi i)}&0&\frac{\kappa}{(2 \pi i)^3}\\ 
\end{array} \right) {\tilde {\underline\Pi}}\ . 
\label{eq:T0}
\ee

Introducing the mirror map $q(z)=\exp(2 \pi i t(z))$ with $t(z)=\varpi_1/(2 \pi i \varpi_0)$, we can write $\underline{\Pi}=\varpi_0(1,t,\partial_t \mathcal{F}(t), 2 \mathcal{F}(t)-t \partial_t   \mathcal{F}(t))$ in terms 
of a prepotential $\mathcal{F}=-\frac{\kappa t^3}{6}-\frac{\sigma t^2}{2}+\frac{c_2}{24} t+\sum_{d=0}^\infty n^d_0 {\rm Li}_3(q^d)$ with $n^0_0=\chi/2$ and $\kappa,c_2,\chi$, which are the topological data of the mirror
to $M$. For general CY $n$-folds the analog of the transformation
in \eqref{eq:T0} is obtained by the $\hat \Gamma$-class formalism 
as explained in \cite{Bonisch:2020qmm}. The main invariant of 
CY $n$-fold motives are the mirror maps $t_i(z_j)$ and the triple couplings in quantum cohomology, also 
known as Yukawa coupling in the string compactification context, as they encode all enumerative invariants.  
For the one-parameter CY three-fold  case, there is only one triple coupling, namely $c_{ttt}(t)=\frac{c(z)}{\varpi_0^2}\left( \frac{\partial z}{\partial t}\right)^3=\partial_t^3 \mathcal{F}$. The last equality is a consequence of GT. With coordinates $x^2=z$ and naive normalization $\kappa=1$, they appear as $Y_k$-invariants and 
structure series $\alpha_k$ in~\cite{Bogner:2013kvr}. For 
our CY \eqref{eq:CY1}, we find explicitly
\begin{equation}
\begin{aligned}
    \alpha_1 &= \frac{1}{(\theta t)(\mathrm x^2)}= 1-64\mathrm x^4-5952\mathrm x^8+\mathcal O(\mathrm x^{12}) \, , \\
    Y_1 &=\frac{c_{ttt}(t(\mathrm x^2))}{16} = 1+32 \mathrm x^4+6944 \mathrm x^8+ \mathcal O(\mathrm x^{12}) \, .
\label{eq:alpha1}
\end{aligned}    
\end{equation}
The recursive definition of the $Y_k$, $k=0,1,\hdots,n-2$ and  $\alpha_k$, $k=0,1,\hdots,n$ 
uses the differential operators ${\cal N}_0=1$, ${\cal N }_1=\frac{\theta}{\varpi_0}$ and ${\cal N}_{k+1}= \theta \frac{1}{{\cal N}_{k}(\varpi_k)}$ and defines $\alpha_k={\cal N}_k(\varpi_k)^{-1}$ and $Y_k=\frac{\alpha_1}{\alpha_{k+1}}$. It is useful to provide the triple coupling in quantum cohomology of CY $n$-folds as in \cite{MR1416356} 
and bring the one-parameter GM in a standard form~\cite{MR1416356,Klemm:1996ts,Mayr:1996sh,Bogner:2013kvr}. 
The theory for multi-parameter GM was worked out 
in \cite{CaboBizet:2014ovf}, see~\cite{Klemm} for review.

For example, for our first CY three-fold  \eqref{eq:CY1} the topological data of the mirror  are  $\chi=-128,\kappa=16, c_2=64$ 
($\sigma=\kappa \ {\rm mod} \ 2=0$) which are precisely those topological data fixing its topological type \cite{MR0215313} and the number of lines $n_0^1=512$ in it, which was observed in~\cite{MR1201748}. The equation \eqref{eq:T0} fixes the integral symplectic basis at $z=0$. The  analytic continuation of ${\underline \Pi}$ to $z=\infty$ is given exactly by a 
Barnes integral representation and to the conifold $z=\frac{1}{2^{10}}$ 
by the construction of a Kuga-Sato variety. The latter implies that 
the transition matrix is given in terms of the periods and quasi-periods of weight four Hecke eigenforms of $S_4(\Gamma_0(8))$.

From \eqref{eq:K3one} we can evaluate the $T^2$ period of the K3 at $x=0$ as residuum, i.e. $\varpi_0^{(0)} = \int_{T^2}\Omega$ with
\be 
\begin{array}{rl} 
\displaystyle{\varpi_0^{(0)}}&=\displaystyle{\frac{1}{(2 \pi i)^2}\oint \oint  \frac{{\rm d} X\wedge 
{\rm d}Z}{\sqrt{Y}}}\\ &=\displaystyle{
\sqrt{z} \left(\sum_{k=0}^\infty \left(-\h\atop k\right) z^k\right)^2=\frac{4}{\pi^2} \sqrt{z} K(z)^2\,.}
\label{eq:varpi0}  
\end{array} 
\ee
This is clearly a solution of ${\cal L}^{(3)}_1$ at $z=x^2=0$ and, in fact, 
one can obtain ${\cal L}^{(3)}_1$ as the third-order differential operator 
that annihilates it.  The other two K3 periods over integral monodromy cycles, which have logarithmic and double logarithmic degenerations at this MUM point, yield precisely those quadratic combinations of elliptic functions that appear as transcendental functions in the 4PM approximation. Similarly, the operator \eqref{eq:apery} can be written as a symmetric square of the second-order operator ${\mathcal L}^{(2)}_2=\theta^2 + 
z^2(\theta+\h)^2 -\h z(\theta^2+34 \theta+5)   $.   

For the second CY three-fold defined by $U^2=P$ with $P$ from 
the Baikov representation of the  graph CY3' in eq. \eqref{eq:P}, we can evaluate, as in \eqref{eq:varpi0}, the $T^3$ period integral in affine coordinates
\begin{equation}
\begin{aligned}
    \varpi^{(1)}_0 &=\frac{8}{(2 \pi i)^4}\oint_{T^4}
                \frac{\sqrt{1+t}}{\sqrt{Q}} \frac{{\rm d} W}{W}\wedge \frac{{\rm d} X}{X}
                \wedge \frac{{\rm d} Y}{Y}\wedge \frac{{\rm d} Z}{Z} \\
                   &=1-\frac{7}{2^8}t^2+\frac{7}{2^8}t^3-
\frac{25711}{2^{20}}t^4+\ldots\, .
\label{eq:toruscy3}
\end{aligned}
\end{equation}
Here, $t=x^2-1$ and the Laurent polynomial $Q$ is $Q=P/(WXYZ)^2$. We can find an operator annihilating the expansion of $\varpi_0^{(1)}$ from an ansatz ${\mathcal L}(\theta,t)$ of order four and six in $\theta$ and $t$, respectively. This operator $\tilde {\mathcal L}^{(4)}$, which is equivalent to the one in \cite{Frellesvig:2023bbf}, has an apparent singularity and lacks the integrality properties of a CY operator. Since it is difficult to 
resolve the $U^2=P$ geometry we reconstruct the CY three-fold motive 
from  \eqref{eq:PF2CY3} by running the $\hat \Gamma$-class argument of 
\cite{Bonisch:2020qmm} backwards. This means we determine an integral symplectic 
basis  by calculating the monodromies around $z=0$ and $z=\frac{1}{2^{10}}$, 
and determine thereby the topological data in \eqref{eq:T0}. By analytic 
continuation we find that such a choice is unique up to an integer $m\in \mathbb{Z}$  with the monodromies $M_{z=0}$ and $M_{z=\frac{1}{2^{10}}}$ ($M^2_{z=\frac{1}{2^{10}}}={\mathbb 1}$)
given by
\be 
\left(\begin{array}{cccc}
1& 1&0&0\\ 
0& 1&0&0\\
2&2m-1&1&-1\\
-4&-2&0&-1\\
\end{array}\right), \  
\left(\begin{array}{cccc}
m& 0&-1&0\\ 
m^2-1&0&-m&0\\
0&1-m^2&0&m\\
\end{array}\right) \, ,
\ee
respectively. This not only determines an integral symplectic basis 
but  also restricts the topological data to $\chi=20 \kappa$, $c_2=\kappa( 6m-5)$ with $\kappa=4$. The genus zero BPS numbers  are integral at both MUM points 
        $\{n^{d}_0\}=\{-640,-27680,-2158729,\ldots, d=1,2,3\ldots\}$ as required~\cite{MR3822913,Bogner:2013kvr}. They appear\footnote{We thank Duco van Straten for pointing this out.} also in case 2.33 of~\cite{MR4668024}. This CY three-fold is defined as Hadamard product~\cite{MR4668024} with  CY operator ${\cal L}^{(4)}_\text{Had} = \theta^4+2^{16} w^2\prod_{k=0}^3(4 \theta+2k+1)-2^4z(4 \theta+1)(4 \theta +3) (32 \theta^2+ 32 \theta+13)$. Its periods are related to the periods $\underline{\Pi}(z)$ of \eqref{eq:PF2CY3} by $\underline{\Pi}(z) = 1/\sqrt{1+2^{10} z}\, \underline{\Pi}_\text{Had}(z/(1+2^{10} z)^2)$, which means that the CY periods associated to the geometry \eqref{eq:P} are also realized in a Hadamard product of two elliptic curves. 

\section{$\epsilon$-factorized differential equation}
\begin{figure}[h]
\centering
\includegraphics[width=0.45\textwidth]{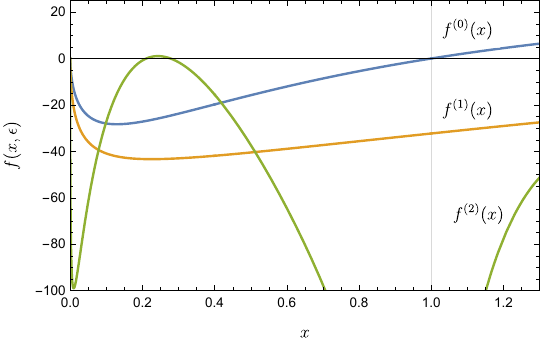}
\caption{Plot of $\epsilon$-orders of $f(x,\epsilon)$ in original variable $x$.}
\label{fig:plots}
\end{figure}

Finally, we want to give explicit results for the $\epsilon$-factorized differential equation in the CY three-fold sector at 5PM-1SF. We do this as an expansion of the rescaled variable $\mathrm x$ to obtain directly integer coefficient expansions. The four new transcendental functions $G_i$ for $i=1,2,3,4$ are given by
\begin{equation}
\begin{aligned}
    G_1(\mathrm x) &\!=\! -6144\mathrm x^4 (1+432 \mathrm x^4+138784 \mathrm x^8 + \mathcal O(\mathrm x^{12}))\, , \\
    G_2(\mathrm x) &\!=\! \frac{128 }{3}\mathrm x^2 (7+2512 \mathrm x^2+29344 \mathrm x^4 + \mathcal O(\mathrm x^{6}))\, , \\
    G_3(\mathrm x) &\!=\! -1536\mathrm x^4 (1+264 \mathrm x^4+66432 \mathrm x^8 + \mathcal O(\mathrm x^{12}))\, , \\
    G_4(\mathrm x) &\!=\! -\frac{64}{3} \mathrm x^2 (7+900 \mathrm x^2-1120 \mathrm x^4 + \mathcal O(\mathrm x^{6})) \, .
\end{aligned}
\label{Gexpansions}
\end{equation}
Notice, that all $G_i(\mathrm x)$ have an integer coefficient expansion after a suitable normalization. The functions $G_1(\mathrm x),G_3(\mathrm x)$ have exponents being only multiples of four whereas $G_2(\mathrm x),G_4(\mathrm x)$ have just even exponents. The final $\epsilon$-form of the connection form $\epsilon A_{\text{CY3}}(\mathrm x)$ in the CY three-fold sector at 5PM-1SF is given by
\begin{equation}
    A_\text{CY3}(\mathrm x) = \begin{pmatrix}
                        K_{11}(\mathrm x) & K_{12}(\mathrm x) & 0 & 0 \\
                        K_{21}(\mathrm x) & K_{22}(\mathrm x) & K_{23}(\mathrm x) & 0 \\
                        K_{31}(\mathrm x) & K_{32}(\mathrm x) & K_{22}(\mathrm x) & K_{12}(\mathrm x) \\
                        K_{41}(\mathrm x) & K_{31}(\mathrm x) & K_{21}(\mathrm x) & K_{11}(\mathrm x)
                    \end{pmatrix} 
\label{epsformCY3}
\end{equation}
with the eight different kernels
\begin{equation}
\begin{aligned}
    K_{11}(\mathrm x) &= -\frac{2}{\mathrm x}+128 \mathrm x+512 \mathrm x^3+32768 \mathrm x^5 + \mathcal O(\mathrm x^7) \, , \\
    K_{12}(\mathrm x) &= \frac{1}{\mathrm x}+64 \mathrm x^3+10048 \mathrm x^7+1878016 \mathrm x^{11} + \mathcal O(\mathrm x^{15}) \, , \\
    K_{21}(\mathrm x) &= \frac{10}{3 \mathrm x}+448 \mathrm x+65408 \mathrm x^3+200704 \mathrm x^5 + \mathcal O(\mathrm x^7) \, , \\
    K_{22}(\mathrm x) &= -\frac{2}{\mathrm x}+128 \mathrm x-2560 \mathrm x^3+32768 \mathrm x^5 + \mathcal O(\mathrm x^7)\, , \\
    K_{23}(\mathrm x) &= \frac{1}{\mathrm x}+96 \mathrm x^3+19040 \mathrm x^7+4199424 \mathrm x^{11} + \mathcal O(\mathrm x^{15})\, , \\
    K_{31}(\mathrm x) &= -5376 \mathrm x-1208320 \mathrm x^3-10149888 \mathrm x^5 + \mathcal O(\mathrm x^7)\, , \\
    K_{32}(\mathrm x) &= \frac{40}{3 \mathrm x}+896 \mathrm x+\frac{438016 \mathrm x^3}{3}+831488 \mathrm x^5 + \mathcal O(\mathrm x^7)\, , \\
    K_{41}(\mathrm x) &= -\frac{476}{9 \mathrm x}+\frac{8960 \mathrm x}{3}+\frac{21856640 \mathrm x^3}{3} + \mathcal O(\mathrm x^5) \, .
\end{aligned} 
\label{kernelsCY3}
\end{equation}
We can see that on the diagonals $A_\text{CY3}(\mathrm x)$ exhibits a symmetry reducing the number of independent kernels which was also observed for the $\epsilon$-form of the banana integrals~\cite{Gorges:2023zgv,Neganew,Pogel:2022yat,Pogel:2022ken,Pogel:2022vat}. Notice, that this is not a general feature for $\epsilon$-deformed differential equations related to CY operators~\cite{Neganew}. With the $\epsilon$-form \eqref{epsformCY3} we can now systematically compute the $\epsilon$-expansion which by analytic continuation can be made global. To demonstrate this we plot in fig.e \ref{fig:plots} the real part of the first three $\epsilon$-orders of the sample function
\begin{equation}
    f(\mathrm x,\epsilon) = f^{(0)}(\mathrm x) + \epsilon f^{(1)}(\mathrm x)+ \epsilon^2 f^{(2)}(\mathrm x) + \mathcal O(\epsilon^3) \, .
\end{equation}
The function $f$ is a linear combination of the maximal cuts of the MIs $I_1,\hdots,I_4$ in the CY3 sector. In particular, we have chosen this linear combination such that $f^{(0)}$ is proportional to the at the conifold vanishing period of the CY three-fold. To be precise we have taken
\begin{equation}
\begin{aligned}
    f^{(0)}(\mathrm x) &= \frac{2}{3} \mathrm x \left(4 \log ^3(\mathrm x)-\pi ^2 \log (\mathrm x)-3 \zeta (3)\right) + \mathcal O(\mathrm x^2)\, , \\
    f^{(1)}(\mathrm x) &= -\frac23\mathrm x \big(8 \log ^4(\mathrm x)+7 \pi ^2 \log ^2(\mathrm x)\, , \\
               &\quad -6 \zeta (3) \log (\mathrm x) \big) + \mathcal O(\mathrm x^2)\, , \\
    f^{(2)}(\mathrm x) &= \frac2{27} \mathrm x (108 \log ^5(\mathrm x)-\left(2700-119 \pi ^2\right) \log ^3(\mathrm x) \\
               &\quad -99\zeta (3) \log ^2(\mathrm x) +4050 \log (\mathrm x)) + \mathcal O(\mathrm x^2) \, .
\end{aligned}
\end{equation}
We can see that at the conifold singularity located at $x=1$ ($\mathrm x=\frac14$) the different $\epsilon$-orders can be smooth or exhibit a singularity. Nevertheless, the analytic continuation can be done beyond this singularity of the differential equation.

\bibliographystyle{JHEP}
\bibliography{CY-letter}

\end{document}